# Supercomputer 3D Digital Twin for User Focused Real-Time Monitoring

Bill Bergeron, Matthew Hubbell, Daniel Mojica, Albert Reuther,
William Arcand, David Bestor, Daniel Burrill, Chansup, Byun, Vijay Gadepally, Michael Houle,
Hayden Jananthan, Michael Jones, Piotr Luszczek, Peter Michaleas, Lauren Milechin,
Julie Mullen Andrew Prout, Antonio Rosa, Charles Yee,
Jeremy Kepner
MIT

*Abstract*—Real-time supercomputing performance analysis is a critical aspect of evaluating and optimizing computational systems in a dynamic user environment. The operation of supercomputers produce vast quantities of analytic data from multiple sources and of varying types so compiling this data in an efficient matter is critical to the process. MIT Lincoln Laboratory Supercomputing Center has been utilizing the Unity 3D game engine to create a Digital Twin of our supercomputing systems for several years to perform system monitoring. Unity offers robust visualization capabilities making it ideal for creating a sophisticated representation of the computational processes. As we scale the systems to include a diversity of resources such as accelerators and the addition of more users, we need to implement new analysis tools for the monitoring system. The workloads in research continuously change, as does the capability of Unity, and this allows us to adapt our monitoring tools to scale and incorporate features enabling efficient replay of system wide events, user isolation, and machine level granularity. Our system fully takes advantage of the modern capabilities of the Unity Engine in a way that intuitively represents the real time workload performed on a supercomputer. It allows HPC system engineers to quickly diagnose usage related errors with its responsive user interface which scales efficiently with large data sets.

## I. INTRODUCTION

The ability to monitor High Performance Computing (HPC) systems in real time and having the ability to forensically diagnose system performance are critical functions for HPC administrators and HPC user support personal [1]. MIT Lincoln Laboratory Supercomputing Center (LLSC) has successfully been using the Unity game engine [2]–[5] for many years to create a Digital Twin [6] of our HPC systems which aggregate and display the large amount of data generated by the various components of the system. These include node hardware, system logs, storage systems, network switches, job schedulers, environmental controls, and accelerators. Utilizing the 3D environment to generate a system view that is representative of the real environment, intuitive to humans, and scalable with the ever growing size and complexity of our HPC systems. With the onset of Artificial Intelligence(AI) [7] and Big Data [8]–[10] systems, it is expected that this trend will continue into the conceivable future.

With the growth of system size and complexity, the need to provide additional ways of analyzing system information has become increasingly clear [11]. Incorporating new tools and perspectives on the data increases the ability to solve issues in real time and to identify factors that cause system performance degradation. The LLSC HPC systems are configured for interactive supercomputing which allows multiple, often hundreds, of users to be running on the system hardware simultaneously [12]. The traditional method of viewing these systems focuses on individual hardware components and sets thresholds for alerts. While effective in identifying system bottlenecks, it tends to be less effective in identifying the root cause of the problem or providing a means of mitigating future performance issues. The need to monitor how individual users interact with the HPC system, and affect other users, in real time, and throughout the duration of their job cycles is a necessity for effective management of an interactive supercomputing environment. Recent work at the LLSC has focused on enhancing the Digital Twin monitoring system to include a user-centric view of system performance, adding accelerator data and visualization, and adding the ability to look through the system history to see how users, or other system issues, impact the individual system components and affect the system as a whole.

## II. HPC SYSTEM MONITORING

System monitoring of large supercomputing environments tends to focus on data collection and rely on enterprise monitoring tools and expert analysts to interpret the data. This often leads to an unproductive work flow as hardware or system configuration errors and inefficient user system usage are not realized in real time, if at all. Problem resolution then requires engaging experts, typically with root level system access, to interpret the data and logs to troubleshoot the system. Solving this workflow issue is a point of focus for the LLSC since the irregular resource consumption of one user can adversely impact the work of everyone on the LLSC systems. This makes proper monitoring and visualization of the system status and user resource consumption critical [5].

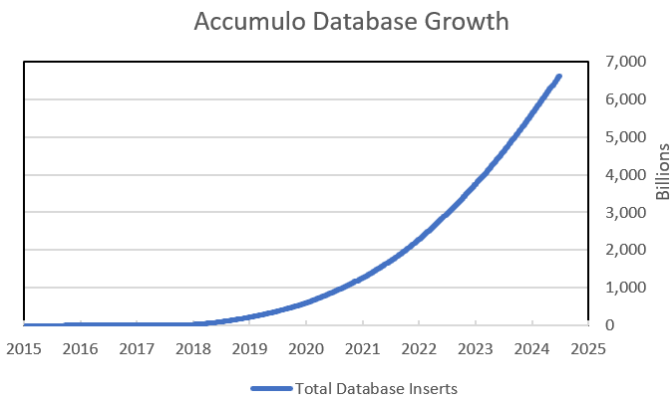

Fig. 1: Data Collection Growth over time for LLSC Accumulo Database

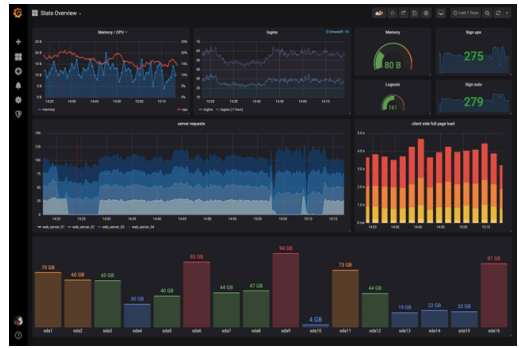

Fig. 2: Grafana Dashboard of a single HPC System Node

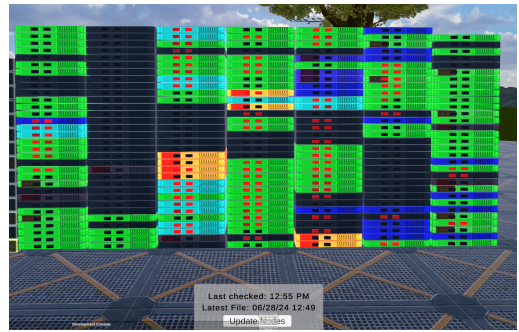

Fig. 3: MandM Digital Twin view of several HPC compute racks under various loads

The focus of many infrastructure management systems is collection and aggregation of system and sensor time-series data. Operating systems, system firmware, HPC storage devices, networking hardware, HPC job schedulers, and environmental systems all generate data continuously and in a variety of formats. The data alone does not unnecessarily equate to increased real time situational awareness as there is a difference between having the data available and having the data in a useful form. Effective visualization tools are needed to interpret the data and give the system support team the ability to analyze and mitigate system issues as they arise. This is where many enterprise tools fall short as they were not designed for the scale of the data collected, and the burden still often falls on human expertise to read and decipher the data with more basic tools and often only after a system issue has become critical.

### A. Data Ingestion and Scaling

The volume of data collected during the operation of an HPC system can overwhelm data aggregation methods, especially as systems continue to scale in size which naturally increases the data collection correspondingly. The LLSC has incorporated our internally developed Dynamic Distributed Dimensional Data Model (D4M) Database [13], [14] utilizing Accumulo [15], [16] for the back-end insertion. This system has proved to be effective and scalable, has grown with the LLSC Monitoring and Management (MandM) Digital Twin, and is still performing effectively after ingesting over seven billion structured data sets from a magnitude higher of ingested raw data across three supporting databases as seen in Figure 1.

Apache Accumulo, combined with D4M, provides a stable, performance-optimized sparse database to store the persistent raw data from the system collectors. Accumulo enables the LLSC to have access to all historical data points collected across the system in a single database and is capable of allowing the LLSC to analyze billions of records for pattern recognition or critical system events and profile them for future early detection.

### B. System Focused Analytics

A major limitation of the most analytic tools, including previous version of the LLSC MandM system, is they focus on the individual hardware components and indicate the "health" of and load on each component. While this is the most obvious way to interpret system performance it often misses the root cause of problematic system issue. Similar to how medical diagnosis often focuses on the outward symptoms of a problem and not the reason for the symptoms. The system approach is most effective when the cause of a problem is failed hardware, but it is far more common for the issue to be related to system usage which is not often obvious when looking at individual system nodes which only indicate symptoms such as load. While the Grafana Dashboard [17] in Figure 2 shows a significant amount of detail about one system component, and Figure 3 shows a detailed view of the condition of multiple system components on the MandM Digital Twin, neither give the bigger picture of how the system is currently performing as a whole. New techniques need to be employed to create a full picture of system usage.

### C. Digital Twinning

In 2011 when the LLSC was exploring new ways to visualize our HPC systems with the volumes of data they produced, we took inspiration from the large scale visual simulations the users were running on our HPC system, TX-Green. The goal was to simulate a complex system of events and derive a visualization that accurately represented how this complex system was interconnected. The goal was to detect faults in the physical system, to provide a better experience for the end user, and enhance the administration of the system. The term for what the LLSC was leveraging was codified in Materials, Structures, Mechanical Systems, and Manufacturing Road Map from the Office of Chief Technologies a year earlier, a Digital Twin. Digital Twin is still loosely defined and continues to evolve; however, it is generally accepted as having "three components; a physical product, a virtual representation of that product, and the bi-directional data connections that feed data from the physical to the virtual representation and information and processes from the virtual representation to the physical" [18], [19]. Over the past decade since describing the practice of utilizing a Digital Twin to enhance the understanding of complex systems, the rise in the use of Digital Twins has become a strategic initiative of organizations and institutions to enhance manufacturing, product development, health care, organizational efficiencies, and general competitive advantage. The market for Digital Twins has even become a cornerstone for nVidia and their creation of the Digital Twin Omniverse [20]. According to McKinsey, the rapid adoption of a Digital Twins approach has created a $48 billion dollar opportunity [21]. Alternatively, the rise in Digital Twins is not just in the commercial space but has also led to an growing body of research on best practices, different methodologies, and novel approaches to using and developing Digital Twins. This is evident from the creation of an annual IEEE conference on Digital Twins with the first proceeding in 2021 [22]. The LLSC is encouraged by the continued development and advances made in the Digital Twin industry and continues to incorporate new features and strategies enhance our own Digital Twin model of our HPC systems using the Unity 3D Engine.

*D. Gaming Engines*

Gaming engines have proven that they can be effective outside of the realm of just game development and can effectively build virtual environments for visualization and training purposes [23]–[25]. Humans are accustomed, and biologically developed, to live in a 3D world and our eyes and brain have evolved to process information in this manner. For example images are processed 6x-600x faster than words [26], subjects perform significantly better using 3D displays [27], and the human brain can process entire images that the eye sees in as little as 13 milliseconds [28], [29]. The 3D gaming environment is uniquely adapted to display vast amounts of information unlike other visual mediums. Studies on 2D vs 3D interfaces indicate more natural ways to visualize hierarchical data should be strongly considered during the interface design process [27]. These observations have led the LLSC to develop the MandM Digital Twin using a gaming engine.

There are two primary game development platforms commonly used in small to midsize projects: Unity 3D and Unreal Engine. Each has a sizeable user base and years of development. Several smaller engines exist such as Godot, Blender Game Engine, GameMaker, Amazon lumberyard, and CryEngine, but none of them are mature enough for our purposes. There are also 3D simulation environments such as nVidia Omniverse which is a suite of various 3D software to tackle various areas of 3D development such as environment, skeletal and facial animation. This software also has the added capability of integrating with many of the previously mentioned 3D apps so any assets can be rendered real-time in Omniverse without needing to export files. A major upside to this is the ability to have real time collaboration in one scene through the Omniverse ecosystem. While this technology is certainly promising for the future of 3D interactive media, it is still quite new, and it is not quite mature enough for our purposes so our decision was between the two major players.

*1) Unreal:* Unreal Engine(UE) is one of the big players in terms of free game engines available for the average consumer and has significant support. Unreal offers much of what is expected from a free engine such as a diverse asset store with everything from 3D assets to scripts, community and official support, documentation, and a modern feature set. Where Unreal has the edge is in graphical development. Unreal is currently and has always been the industry leader in cutting edge graphics technology with UE5 even being used to make digital television and movie sets. The engine also has a proven track record being used in Digital Twinning applications such as prototyping for wind power [30]. It offers intuitive networking and visual scripting solutions as well the ability to assist designers in getting accustomed to the engine. Overall, Unreal is a very powerful engine with widespread use across multiple industries and has support for many applications.

*2) Unity:* Unity has a robust user community because it has been the most popular independent game engine for over a decade. The level of community support and user generated documentation is currently unmatched in the game development scene. Unity also allows for fast iteration, quickly switching between the editor and play mode, making the quick construction of demo versions a hassle-free process. Unity makes porting code to various platforms easy, coming with support for Windows, Mac, IOS, and Linux right out of the box. This is extremely important at the LLSC where a variety of operating systems are used. While Unity is not used for digital movie sets it is still a contender in the Digital Twinning sphere. It has been used for educational purposed such as simulating an assembly line or warehouse stackers [31]. The engine also comes with a diverse asset store that offers a range of assets from scripts, 2D art, 3D models, audio, and multitude of other tools to develop a project. These tools are already integrated with the Unity Engine, and after acquisition they can be easily imported with the built-in package manager. It also has the option for more advanced features such as its HD Render Pipeline and the Entity Component System , all while being free for project development.

*3) Selection:* Unity was chosen for the MandM Digital Twin because it provides an appreciable balance between ease of use for smaller projects and sophisticated tools when scaling up a project's complexity. The main downside with Unreal for MandM is that the user experience for Unreal is not as easy to learn for small teams making a modest project. Unreal has a higher learning curve and it's C++ integration is trickier to learn than Unity's C#. The main disadvantage to using Unreal engine for smaller projects is the speed of iteration. It takes the engine 10-20 seconds to hot-reload after making changes to the code which starts to eat up development time with every change made. For larger studios this is less of an issue because they have more powerful computers, and the benefit of higher graphical fidelity outweighs the downside of spending slightly more on project load times. While Unreal offers better features for high fidelity commercial games, Unity makes more sense for a small project like MandM where graphical fidelity is not the primary goal. Overall, the Unity game engine was the clear best choice for this project with all the features it offers such as being quick to use, having an endless amount of community support, the option for more advanced features.

## III. PREVIOUS APPROACH

The MandM Digital Twin has effectively managed to collect and display the vast quantity and variety of data gathered by the operation of the HPC systems. Utilizing the visual and intuitive advantage of an interactive 3D gaming environment, overall system visualization in real time has been achieved. System components with visual indications of load, performance, and hardware status are displayed together as a complete system with the ability to drill down into sub-components to gather additional information and to troubleshoot system activity. But the approach was still largely relying on the system components to indicate status, and it required significant time searching to find root causes for displayed problem. While user activity was contained within the different system nodes there was not an effective way to look at overall user activity and the effects on the larger system.

### A. Interactive Supercomputing

One aspect of the LLSC HPC environment which heavily influences our methodology is our focus on enabling Interactive Supercomputing. This method of system use allows for multiple users to be interactively utilizing the system simultaneously which results in a wide variation in usage workflows. High utilization of different system components is common. High CPU, GPU, storage I/O, Network I/O, and/or system memory utilization can occur using these various workflows. While user system limits are established it is common for users to maximize their resources often by non-optimized usage, and in combination with other users can affect the overall system in negative ways. This method of operation requires that HPC support personnel are aware of the various users activity on the system. When the HPC systems and the active userbases were smaller, this was a manageable human task, but as systems have scaled for AI, Big Data, and Large Language Model (LLM) usage along with a sizeable increase in the userbase, other analytical methods are required to manage systems with large userbases.

### B. Data to Information and Analysis

Another significant piece of the MandM system is data conditioning and ingestion. As mentioned previously, we have used D4M and Accumulo to great effect to create a scalable data gathering infrastructure, but it is equally important to have relevant data to analyze, turning the raw information into useful data. This is accomplished by deconstructing the manual process performed by HPC system experts and mining the data gathered that tends to relate to poor user or system performance and job failures. Then continually track and record this data to attempt to identify patterns or conditions that typically lead to the performance degradation and system bottlenecks. Figure 4 shows the flow from raw data to processed data sets to visualization as a Digital Twin. The development of a structured data set which can then be queried, displayed, alerted against, and correlated to user jobs running on the system is a vital part of the MandM process. The last crucial feature of the collected data infrastructure is flexibility. The ability to change and expand the data included and adjust the alert threshold levels and methods of observation is an integral part of the architecture used in developing the LLSC data set.

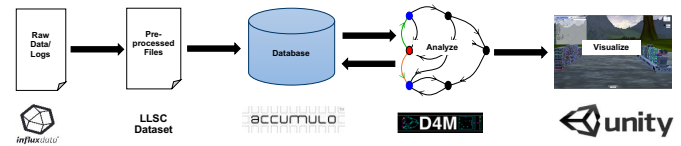

Fig. 4: *Data Flow from raw data gathering, to ordered data set, to processed data, and to Digital Twin*

## IV. RESULTS

### A. Enhanced Situational Awareness

The overall goal of the MandM Digital Twin is to create greater situational awareness for the LLSC support team and to provide a stable, reliable, and robust system for the end users. The backend data gathering techniques to generate useful information from the raw data, the D4M/Accumulo data storage and retrieval infrastructure, and the use of the Unity Game Engine to create a Digital Twin environment have all helped accomplish this task. The goal going forward was to build upon these same techniques with tools to enhance our awareness further. Two possibilities were identified to accomplish this: the use of a user-centric view of the data and adding in a way to access data history from the 3D environment, both seen in Figure 5. The 3D Virtual User system creates another level of awareness for the operator of this system because they can see "problematic" users standing out from the other users on the system, making for quicker identification of system issues. The History Loading System enables the operator to leave the real-time view of the system and to queue up diagnostic data on the entire cluster going back for several hours or more and parse through and visualize it. It allows the operator to get a better picture of the resources used over time for any user on the cluster, see the full job cycle of each individual user's job, and that job's impact on the overall HPC performance for the duration of its runtime. These two techniques we believe have markedly increased the overall situational awareness of the LLSC utilizing MandM.

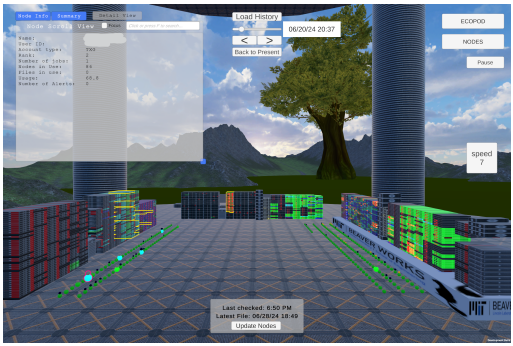

Fig. 5: Full System View including Virtual User Avatars and Data History

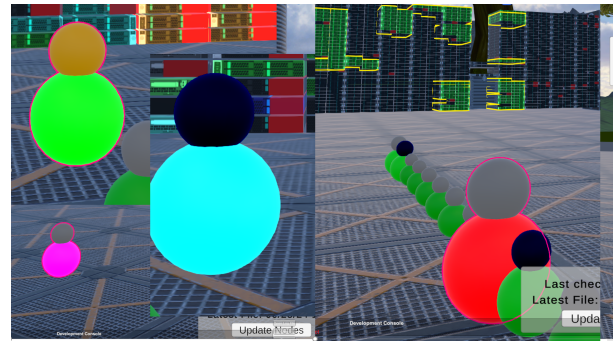

Fig. 6: Virtual User Avatars Indicating Workload and System Error Correlation

### B. Performance

Performance has been one of the biggest challenges in the development of the 3D monitoring environment. The number of objects associated with the HPC systems and the amount of data being process in real time for each object is a major factor in system performance. If the gaming environment becomes "glitchy" or there is prolonged buffering during use, the tool quickly becomes unusable for system troubleshooting. So to allow the MandM system to add new features such as the User View and especially a performance heavy feature like the History System required some extra headroom to run properly. The first significant change made was to make all the GameObjects static in the scene. Static GameObjects require fewer batching calls on the GPU and since there are well over one thousand objects being tracked this is significant. Secondly, we found code issues that caused the GameObjects to duplicate spawning, which led to not only doubling the GameObjects in the scene, but caused the colliders on those nodes to force a virtual physics collision every frame. The obvious solution was to stop the double nodes spawning, but also to switch all colliders to being triggers which prevented physics calculations on the GameObjects. This way the GameObjects would only register by being clicked by the mouse and therefore greatly reduce the processing being done by MandM which greatly improved performance.

### C. User Focused Analysis and Visualization

When the MandM Digital Twin was originally developed, its focus was to identify system bottlenecks on the individual compute nodes and other hardware components. While this proved to be useful information it often fell short in identifying the root cause of overall system performance slowdowns, which was typically system users running in non-optimal configurations. Creating a 3D Virtual User Avatar (VUA) for each user and consolidating all data related to the individual system users was our solution to this problem. The VUAs display is game adjacent to the supercomputing cluster that they are running on. The VUA, by default, is small and green when running withing the normal parameters of the system, but as they increase the number of nodes and files they access, the UVA appearance changes correspondingly. The appearance parameter is determined by our algorithm of "Usage" which is calculated using 80% of the number of nodes and 20% of the number of files in use with the Usage range from 0-100 as seen in Figure 6. When the usage is higher, the 3D UVA size will increase and it will also change color. The colors are green for normal usage, cyan to indicate a user utilizing significant resources, then red for users who are adversely affecting system performance. This gives a single overall system view where both high load systems and high load users are viewable and allows for drilling down into either (or both) to obtain more information and to see how the users and system component usage correspond. For example, When clicking on a VUA it will highlight the user and all nodes that they are currently running on. The opposite is also true, when clicking on a node it will highlight the node and the users running jobs on it. When a user is selected all their relevant information is displayed as well, that being their name, ID, rank, number of nodes in use, files in use, number of jobs running, alerts, and Usage.

This VUA addition makes it significantly easier to locate root issues of poor user performance or a system slowdown on the HPC system. When a user is using resources in a non optimal way, the VUA appears large and bright red, instantly standing out from the other users running on the system. This allows the operator the ability to see where the system is being affected with one click on the problem VUA to see all the resources they are currently using. The VUA also works seamlessly with the History system as described in the next section.

### D. Incorporating Data History

One aspect missing from previous iterations of the MandM Digital Twin was a way to see the system usage over time. While the priority is seeing the system in real time, it is often useful to see how a system node, user, or some combination was affected during the near history of the system and see how a problem developed. The challenge was how to do this is a way that enhanced the ability to troubleshoot a problem and not be cumbersome to perform. The solution was to load into the game memory recently collected data and to be able to step through each data segment that corresponded to a moment in time. Currently data segment files are generated roughly every 5 minutes. When loading the node data history, the thought process started with how to approach loading all the data files generated by the cluster in a way that would be both fast and easy to parse through. One option would be to store the files on the cluster then load each file in a case-by-case basis where the user would have to request a specific file, then load that file off the network to see it in the simulation. That system would be fast to switch the view mode, but slow to load each individual file. The second option would be to store the files on the cluster, load all the files corresponding to a certain time interval, and store the data locally into one array. From there it would be extremely fast to parse from file to file. There would be a lengthy initial load time, but once loaded it would be quick to use. Another option would be to store the data history files locally on the machine running the simulation which would allow for an almost instantaneous loading of all the data files. The downside here is that a significant portion of storage space would have to be dedicated to storing these files locally which would also be more difficult to scale up if we were to increase the amount of data being stored to a week, or a month, or more.

The approach that was taken was the second, which is storing the files on the HPC system being monitored, caching its data locally as an array of strings, and parsing through the array to display the data. The data is stored in text comma separated value (CSV) files similar to the approach taken to load the spatiotemporal data of cities into Unity by Helbig, et al. [32]. The reason why this approach was chosen over the others is because it provides the best balance of user experience and offloads the intensive resource usage onto the cluster, allowing the system to scale up better. It's fast enough to load all the data at once and then cache it locally. Much of the initial load time comes from the network speed, not the front-end processing. The loading takes about 0.13 sec/file so for 24 hours of data which involves 288 files, the load time was 37.44s. When the files are stored locally, loading is almost instantaneous, but it is only when the files are loaded from the cluster that the loading slows, with remote access through a Virtual Private Network (VPN) being the slowest; see Figure 7. When loading the data onto the nodes from the array the framework was already set up to load when it updates every few minutes. What was needed was a system to call the functions which loads the data onto the nodes and to find an intuitive way for the user to parse through the data array.

If the loading were to happen on a single thread, it would lock up the game for the duration of loading which would not be ideal for the user experience. It would be more convenient if the loading happened on a different thread. Unfortunately, Unity's engine code does not support threading but we were able to get around that. The data can not be loaded directly into any Unity related resources, but it can be loaded into a C# array that Unity is able to access. So, using the C# Task system we were able to put the loading process on a different thread allowing the users to still interact with the system while the loading was ongoing. It also allowed for visual feedback on the percentage of files being loaded so the user is aware of how much loading time is left.

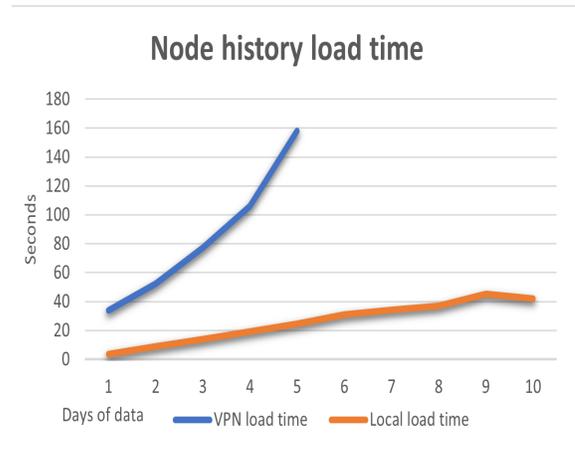

Fig. 7: History File Load Time on local network and through VPN

The History system is a novel approach to presenting system operators with enhanced awareness to quickly identify areas of system performance degradation on the cluster over time. It accomplishes this by allowing the user to quickly load cluster history data locally and parse through it with ease which allows the operator to easily monitor how specific users and the jobs they run affect the cluster in real time and throughout their job cycles. This approach to monitoring a user's impact on the cluster is also scalable with the ever-growing size and complexity of the systems with minimal performance impact on the monitoring software.

*E. Accelerators View*

As accelerators become increasingly common and important in modern HPC systems, it's important to shift the focus of MandM to prioritizing hardware accelerated nodes. We have implemented a simple indicator for GPU units on these nodes, as shown in Figure 8, indicated by two small cubes that vary in color based on the load on each GPU component: black when not in use, then increasingly brighter shades of red as the load increases. This is a useful indicator of which nodes have accelerators and how much those are in use without having to click on the detail view; however, there is certainly room for expansion in the future. Nodes with accelerators could have more weight in the User's usage or there could be a focus view specifically for accelerated nodes.

*F. Conclusion and Future Work*

*1) Future work:* This latest iteration of the LLSC MandM Digital Twin with the added features of VUAs, History, and

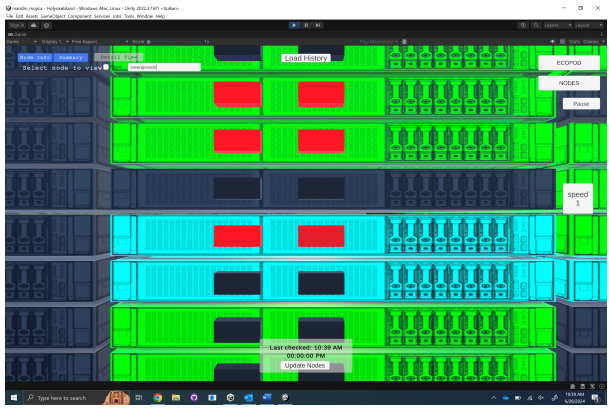

Fig. 8: Nodes with GPU load indication

accelerators has increased situational awareness while maintaining performance as the data scale has increased but there is still plenty of opportunity for improvements. A significant increase in performance to MandM could be achieved in the future by incorporating Unity's Entity Component System (ECS) [33]. In typical game engine design, the approach is to have multiple instances of a type of object with its own unique copy of its data and behavior. ECS takes a different approach to this which separates the Entity (the object), the Components (the data), and the system (what that data does). The main benefit of this approach to object instancing is that instead of many data heavy copies of objects all with their own unique copies of behavior systems, there is only one instance of a system that parses through and operates on every entity. An example would be a collection of car entities all with a unique "speed" component which only contains the speed data. In every frame, the singular speed system would parse through every entity with a speed component and move it forward according to it's current speed. This lightweight approach to object instancing would be ideal for MandM's situation with thousands of objects in the scene at once.

The History system could also be expanded to other related systems being monitored including the environmental systems of the "EcoPod" enclosures used as data centers for the LLSC systems. It would echo a similar process to the nodes themselves with the data files being stored on the cluster and then being loaded into Unity when entering History mode. Similar to the detail node view, the same slider would be used to parse through the data array, but a different user interface (UI) would be used that fits the EcoPods more effectively.

Another method of increased efficiency could be accomplished in the real-time and historical data loading process by replacing CSV files with a serialized file type, especially with the expectation of future growth of the HPC systems. These serialized files could contain structs and other simple data types which would allow the loading process to be faster and simpler than parsing through strings. It would also be compatible with the Unity Job system [34] which is significantly lighter weight than the C# Task system.

ACKNOWLEDGMENTS

The authors wish to acknowledge the following individuals for their contributions and support: LaToya Anderson, Bob Bond, Alex Bonn, Alan Edelman, Jeff Gottschalk, Chris Hill, Charles Leiserson, Kirsten Malvey, Heidi Perry, Steve Rejto, Mark Sherman, Marc Zissman.